\documentclass[11pt,twoside]{article}
\usepackage{asp2010}

\resetcounters

\begin{document}

\title{The AGB population of NGC 6822 }
\author{L. F. Sibbons$^1$,  M.-R. L. Cioni$^1$, M.Irwin$^2$ and M. Rejkuba$^3$}
\affil{$^1$ University of Hertfordshire, Science and Technology
            Research Institute, Hatfield AL10 9AB, United Kingdom}
\affil{$^2$  Institute of Astronomy, Madingley Road, Cambridge,
  Cambridgeshire, CB3 OHA, United Kingdom} 
\affil{$^3$ European Southern Observatory, Karl-Schwarzchildstr-Strasse 2,
  D-85748, Garching bei M\"unchen, Germany}

\begin{abstract}
The metallicity gradient and the stellar distribution within the Local
Group dwarf galaxy NGC 6822 has been studied photometrically 
using asymptotic branch stars (AGB). In order to study the stellar and
metallicity distribution, the carbon- and oxygen-rich AGB  
stars have been isolated using deep high-quality near-infrared UKIRT
photometry. The ratio between them, the C/M
ratio, has been used to derive the [Fe/H] abundance within the
galaxy. The [Fe/H] abundance and stellar distribution were analysed as
a function of galactic radius. A mean C/M 
ratio of $0.288\pm0.014$ has been found which corresponds to an iron abundance of [Fe/H] 
$=-1.14\pm0.08$ dex, with variations in the north and
south, as well as at larger galactocentric distances. Variations in the
magnitude of the tip of the red giant branch has also been
detected.
\end{abstract}

\section{Introduction}
\label{intro}
The Local Group dwarf galaxy NGC 6822 is similar in size, structure and
metallicity to the Small Magellanic Cloud (SMC). The detection of RR
Lyrae stars by \citet{2004ASPC..310...91B} shows that NGC 6822 began
forming stars at least 10 Gyrs ago. Observations of the many HII and OB
associations confirm that star formation is still ongoing. Young blue
stars trace active star formation in both the bar and the large HI envelope 
\citep{2006AJ....131..343D,2003ApJ...590L..17K}. Old and intermediate age red giant branch 
(RGB) and AGB stars, have been detected in the bar and the HI envelope of
the galaxy \citep{2005A&A...429..837C} but according to \citet{2002AJ....123..832L} are concentrated in a
stellar halo. Due to its close proximity ($\sim490$kpc) \citep{1998ARA&A..36..435M} NGC 6822 has been the focus of numerous
investigations of its stellar content, e.g. \citet{2006AJ....131..343D}. HII  
regions and planetary nebulae have also been studied extensively. Several estimates have been made of the iron
abundance across a range of ages. \citet{1996AJ....112.1928G} obtained
[Fe/H] $=-1.5 \pm 0.03$ dex from the slope of the RGB. Further analysis of
RGB stars by \citet{2001MNRAS.327..918T} yielded a value of [Fe/H]
$=-1.0 \pm 0.3$ dex from the strength of CaII absorption
lines. Looking at the younger stellar population, 
\citet{2001ApJ...547..765V} derived an 
average value of [Fe/H] $=-0.49 \pm 0.22$ dex from optical
spectroscopy of A-type super-giants. These results suggest, as expected,
that the chemical enrichment of the interstellar medium has been
a continual process due to multiple stellar generations.

\section{Observations and data reduction}
\label{obs}
Observations were obtained using the Wide Field Camera (WFCAM) on the
$3.8$m United Kingdom Infrared Telescope (UKIRT) in Hawaii during two
runs, in April $2005$ and November $2006$. WFCAM is comprised of four Rockwell-Hawaii-II
infrared detector arrays (HgCdTe 2048 $\times$ 2048), with a combined
field of view of 3 deg${^2}$ (a tile) on the sky and
a pixel scale of $0.4^{\prime\prime}$. A mosaic of four tiles was
obtained in three broad-band filters ($J$, $H$ and $K$)
covering an area of $1$ deg$^2$ centred on NGC 6822 ($\alpha=19^h 44^m 56.0^s, \delta=-14^{\circ} 48^{\prime}
06^{\prime\prime}$). The total integration time of observations in the $J$
band was $150$ sec, from the co-addition of three $10$ sec
exposures for each pointing in a dithered pattern of $5$ points. In the $H$ and $K$ band the
total integration time was $300$ sec for $6$ exposures following a 
similar serving strategy. Data reduction, including all the standard steps for
instrumental signature removal was completed using
the WFCAM pipeline at the Institute of Astronomy in Cambridge. Sources
extracted using the pipeline were given a morphological classification
from which assorted quality control measures are computed. Astrometric
and photometric calibrations were then performed based on 
the 2MASS point source catalogue
\citep{2004SPIE.5493..411I}. Duplicated sources were removed
using the photometric error and the morphological classification to
select a `best' unique entry per object and produce a final catalogue
($\sim 375,000$ sources). With the exception of one frame observed
under poor conditions, the average seeing across the two observing
runs was between $\sim 0.9-1.1^{\prime\prime}$. Observations reached a depth of
$20.61$, $20.03$ and $19.64$ mag in the $J$, $H$ and 
$K$ bands respectively. The AGB sources are expected to have
magnitudes brighter than $K=17-18$ mag, and therefore our photometry
is almost $100\%$ complete at these magnitudes. Reddening values
across 
NGC 6822 have been found to vary widely from the centre to the outer regions
\citep{2009A&A...505.1027H}. This variation is
likely to be intrinsic to NGC 6822 rather than a result of variation in the foreground 
component. As there is no published extinction map for this galaxy no
corrections were made for internal reddening. A
reddening value of $E(B-V)=0.25$ mag has been applied to account for
foreground extinction \citep{1998ApJ...500..525S}. To ensure a reliable data set, only 
sources that were identified as stellar (or probably stellar)
and that were consistently detected in all three photometric bands,
have been selected for the analysis conducted here.

\section{Foreground removal}
\label{foreground}
The observations suffer from heavy foreground
contamination from stars in the Milky Way (MW).   
This was removed using a colour 
selection criterion based on the work of
\citet{2008MNRAS.388.1185G}. The observed area was subdivided into a
grid of $100$ regions, each of dimensions $6\times6$ arcmin, roughly centred on the
optical co-ordinates of the galaxy. Sources from the 
centre of the galaxy were plotted on a colour-colour diagram ($H-K$,
$J-H$), and then overlaid with sources from a region at the periphery
of the galaxy that was dominated by MW foreground stars. There is a clear
separation in the colour between the sources from the centre and those
from the outer region at $(J-H)=0.72$ mag, with few MW stars redder
than this. Assuming that foreground 
stars are evenly distributed across the surface of the galaxy the stars with
$(J-H)>0.72$ mag are most likely members of NGC 6822.

\section{The tip of the RGB}
\label{TRGB}
\begin{figure}[ht]
\begin{minipage}[t]{0.5\linewidth}
\centering
\includegraphics[scale=0.3]{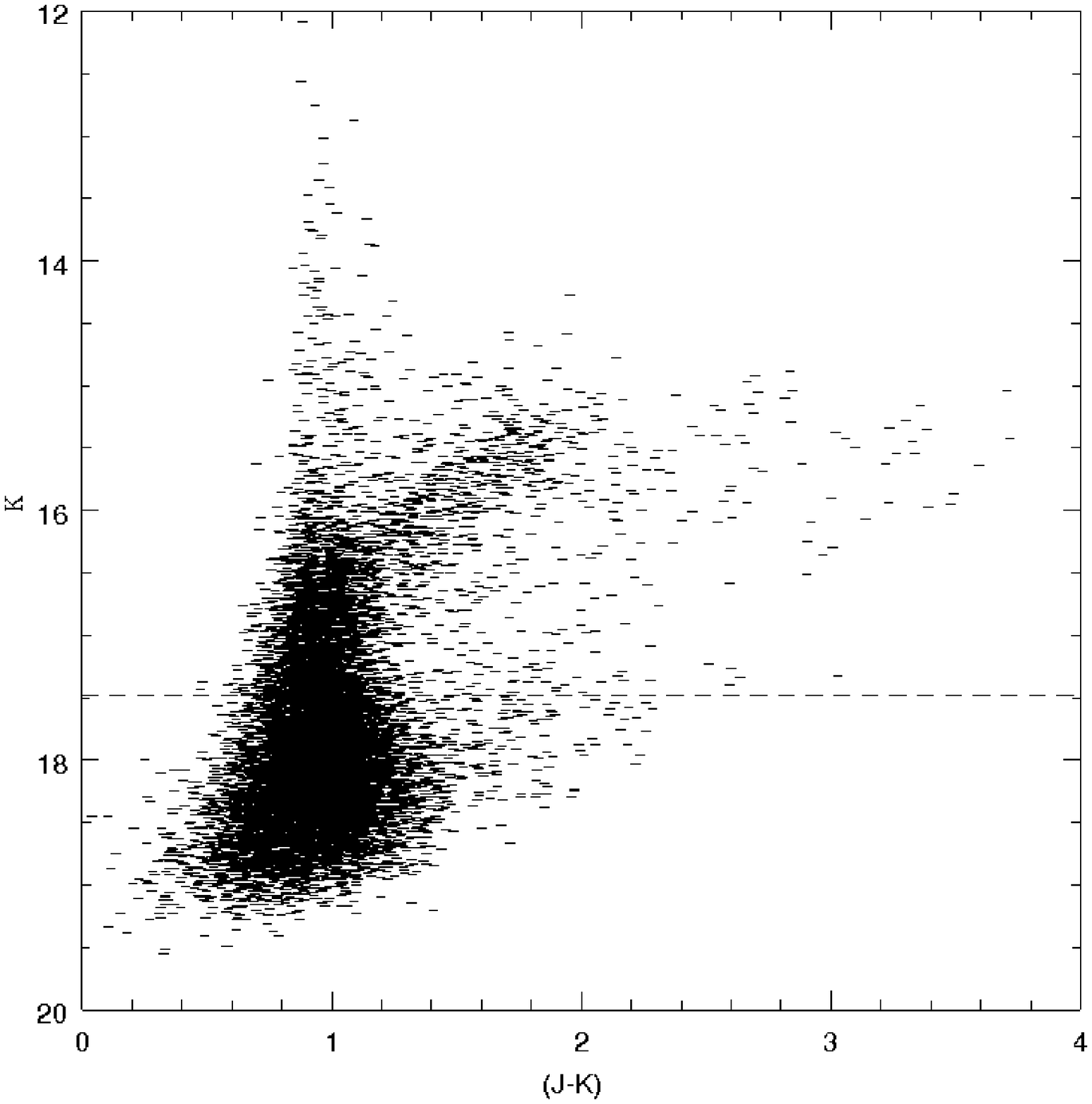}
\caption{CMD of NGC 6822 sources. The dashed horizontal line marks the
average position of the TRGB $K=17.48$ mag. }
\label{postforecmd}
\end{minipage}
\hspace{0.4cm}
\begin{minipage}[t]{0.5\linewidth}
\centering
\includegraphics[scale=0.3]{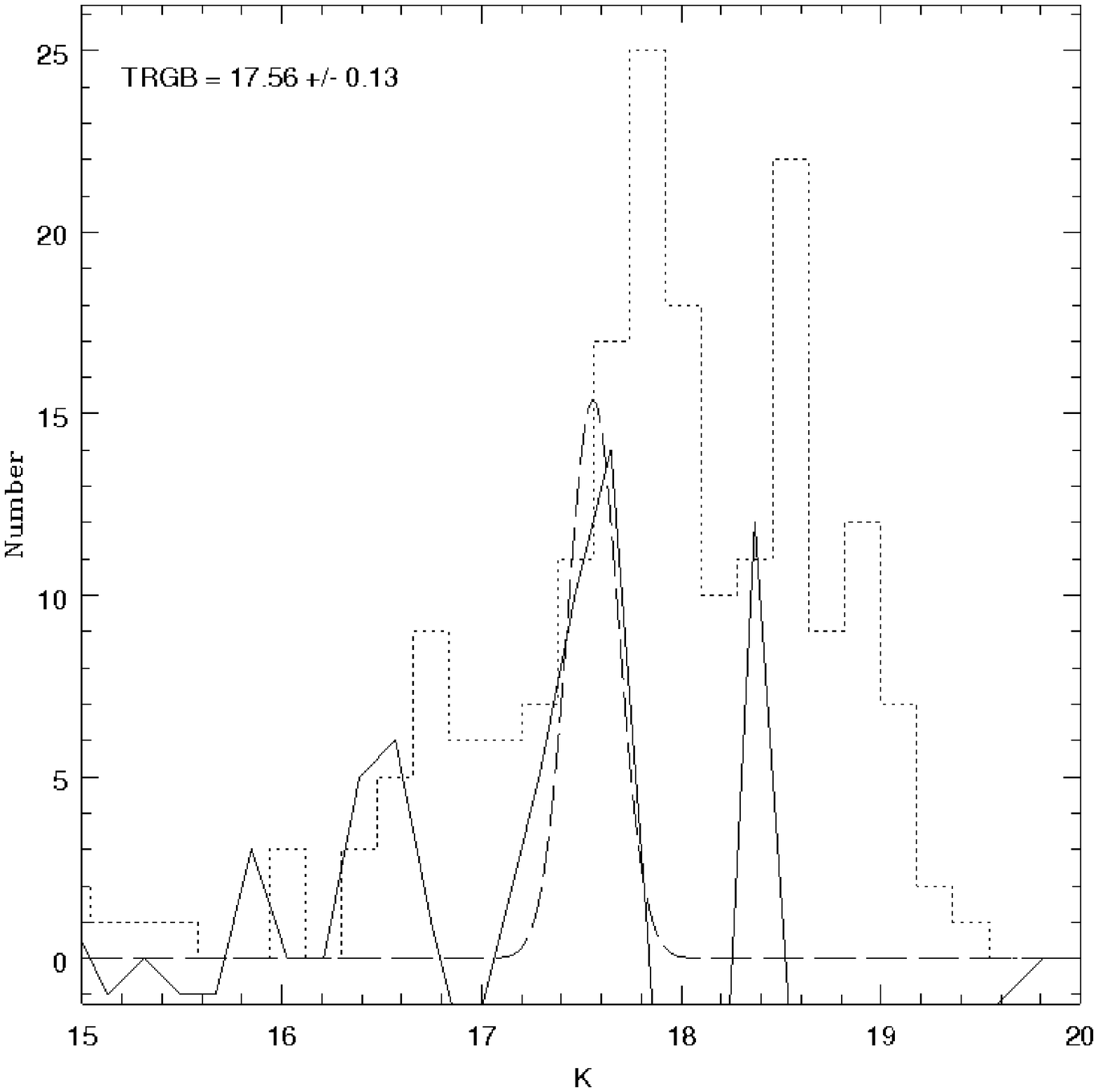}
\caption{$K$ magnitude distribution (dotted line), Sobel filter (solid
  line) and Gaussian fit (dashed line).}  
\label{sobel}
\end{minipage}
\end{figure}

The luminosity discontinuity at the tip of the red giant
branch (TRGB) is one of the most prominent
features in the magnitude distribution of a stellar population and
is commonly used to identify AGB stars in galaxies outside the MW,
e.g. \citep{2006A&A...454..717K}. Fig. \ref{postforecmd}
shows a colour-magnitude diagram (CMD) of sources belonging to NGC 6822. A decline in the
number of sources at bright magnitudes identifies the TRGB at
$K\sim17.25$ mag. In order to validate the position of the TRGB, magnitude distributions were
constructed for each region of the grid and the TRGB discontinuity was
identified first by eye. The TRGB was found to vary across NGC 6822 by $\Delta K=0.96$ mag,
with an average value of $K=17.29\pm0.03$ mag, in reasonable agreement with
the value estimated from the CMD. As a further check, the Sobel edge detection algorithm was also
applied to each magnitude distribution to confirm the position of the
TRGB discontinuity \citep{1993ApJ...417..553L}. The
filter effectively measures the slope of the histogram, and the first
derivative produces a peak where the slope of the
original histogram is greatest. After the peak corresponding to the
TRGB was identified with the Sobel filter, it was fitted with a
Gaussian, whose mean and sigma were taken as the TRGB 
magnitude and associated error (Fig. \ref{sobel}). Corrections
have been applied to the derived TRGB magnitudes, to account for the
slight bias in the Sobel filter, according to the work of
\citet{2000A&A...359..601C}. Using the Sobel filter the TRGB was identified
in each region of the observed area and a variation of $\Delta K=1.36$
mag was found across the surface of the galaxy, 
with an average value of $K=17.48\pm0.26$ mag.
For the purpose of identifying the AGB stars in the sample a single
TRGB value of $K=17.41\pm0.11 $ mag will be used. This value was determined
from the grid region with the highest concentration of sources ($\sim
2160$). Note, the size of the TRGB magnitude variation indicate that the
effects of metallicity, age and reddening variations within the NGC 6822 are not
negligible.    

\section{The AGB stars}
\label{jk}
\begin{figure} 
\centering
\includegraphics[scale = 0.3]{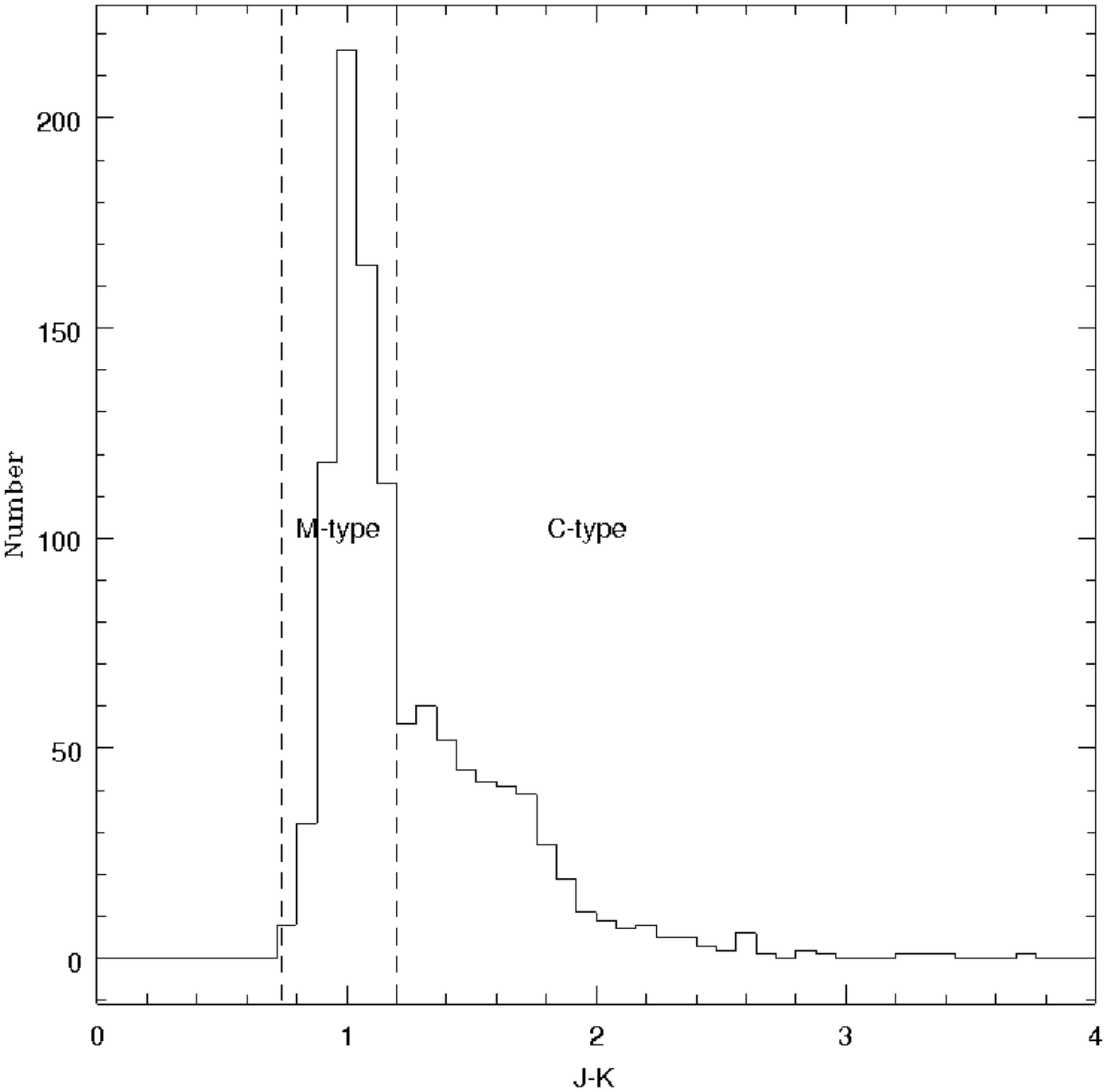}
\caption[]{\footnotesize{Colour histogram of sources from the centre of the
    observed area with $K<17.41$ mag. Vertical dashed lines mark the blue limit at
    $(J-K)>0.74$ mag and the colour separation at $(J-K)=1.2$ mag.}}  
\label{finhist}
\end{figure}

There are two spectral types of AGB stars, M-type (C/O $<$ 1) and C-type (C/O $>$ 1),
respectively. These two spectral types are clearly separated on a CMD
($J-K, K$). M-type stars follow a vertical
feature above the TRGB with a large range of magnitudes and nearly
constant colour, while C-type stars display a smaller range of
magnitudes but a wider range of colours,
resulting in a `red tail' extending diagonally upwards on the CMD (Fig.\ref{postforecmd}). Thus, colour separation has commonly been used
to identify the C- and M-type components of the AGB 
population \citep{2005A&A...437...61K}. The best
estimate of the position of this colour separation from the CMD (Fig. \ref{postforecmd}) falls at
$(J-K)\sim1.1-1.2$ mag. The exact position has been verified by eye from the discontinuity in
the colour histograms of AGB sources (Fig. \ref{finhist}). In this
figure, the highest peak relates to M-type stars 
followed by a significant drop and then a plateau which contains the
C-type stars.   

The position of the colour separation varies across the surface
of the galaxy by $\Delta(J-K) = 0.2$ mag. This may be due to
extinction or variations in the stellar population (metallicity and
age). As for the TRGB magnitude, a standard value ($(J-K)=1.2 \pm
0.1$ mag) was selected for the purposes of separating C- and M-type 
stars in the sample. The application of a sharp colour selection criteria suggests a strict
transition between the two evolutionary types of AGBs, however this is
unlikely to be the case in reality. After selecting the AGB
stars from the sample, they span a colour range of $0.62<(J-K)<4.08$ mag. Therefore, in
accordance with \citet{1988PASP..100.1134B}, it was
decided to apply a `blue limit',$(J-K)=0.74 \pm 0.01$ mag, to the
selection of M-type stars in 
order to exclude late type K-stars, which most probably are not AGB
stars.

\section{The C/M ratio}
\label{ratio}
A catalogue of 3820 AGB sources has been generated from this work,
containing $854$ C-type stars and $2966$ M-type
stars. Spatial variations in the C/M ratio are a
useful indirect indicator of the metallicity distribution, where a
higher ratio implies a lower metallicity.  The calibration of
the C/M-[Fe/H] relation has been refined recently by
\citet{2009A&A...506.1137C}, based on the work of
\citet{2005A&A...434..657B}. Using this relation a mean C/M
ratio of $0.288 \pm 0.014$ has been determined for NGC 6822,
which corresponds to a mean [Fe/H] value of $-1.14 \pm 0.08$ dex. This
is consistent with the mean metallicity of the old RGB population in
this galaxy determined spectroscopically by
\citet{2001MNRAS.327..918T}.

\section{Conclusions}
\label{con}
Here, we have derived an overall C/M ratio of $0.288 \pm
0.014$ and an iron abundance of $-1.14 \pm 0.08$ dex. Examining the
ratio and iron abundance in different regions of the grid used above we
see a spread of $\Delta \mathrm{(C/M)} = 1.33$ and $\Delta \mathrm{[Fe/H]}=1.45$ dex
across NGC 6822. The remaining analysis of this data will focus on the distribution of the
AGB sources in NGC 6822 using source density plots and the variation of
the C/M ratio and TRGB magnitude with galactocentric distance and
azimuthal angle. These results are still under examination but it is
clear that NGC 6822 has and will continue to be an interesting and
challenging object to study.

\bibliography{sibbons}

\end{document}